\documentclass[sigconf]{acmart}
\usepackage{amsmath,amsfonts}

\usepackage{algorithmic}
\usepackage{graphicx}
\usepackage{textcomp}
\usepackage{xcolor}
\usepackage{dblfloatfix}
\usepackage{tabularx}
\usepackage{cuted}
\usepackage{multicol} 
\usepackage{caption}
\usepackage{subcaption}
\usepackage[shortlabels]{enumitem}
\usepackage{multirow}
\usepackage{pbox}
\usepackage{xspace}
\usepackage{makecell}

\newcommand{\eg}{\textit{e.g.}\@\xspace}
\newcommand{\ie}{\textit{i.e.}\@\xspace}
\newcommand{\etal}{\textit{et al.}}
\newcommand\invertedComma[1]{``#1''}
\newcommand\lineQuote[1]{``\textit{#1}''}

\usepackage{booktabs}
\usepackage{array}
\usepackage{colortbl}
\usepackage{caption}

\AtBeginDocument{%
  }

\setcopyright{acmlicensed}
\copyrightyear{2024}
\acmYear{2024}
\acmDOI{3613904.3642692}
\acmConference[CHI '24]{ACM (Association of Computing Machinery) CHI conference on Human Factors in Computing Systems}{May 11--16, 2024}{Honolulu, HI, USA}
\acmPrice{15.00}
\acmISBN{979-8-4007-0330-0/24/05}

\usepackage{acmart-taps}
\aptLtoX[graphic=no,type=html]{}{\def\toprule{\specialrule{.2em}{.1em}{.1em}}}

\begin{document}
\title{Exploring AI Problem Formulation with Children via Teachable Machines}

\author{Utkarsh Dwivedi, Salma Elsayed-Ali, Elizabeth Bonsignore, and Hernisa Kacorri}
\email{{udwivedi, sea, ebonsign, hernisa}@umd.edu}
\affiliation{%
  \institution{University of Maryland}
  \city{College Park}
  \state{Maryland}
  \country{USA}
}





\begin{abstract}

Emphasizing problem formulation in AI literacy activities with children is vital, yet we lack empirical studies on their structure and affordances. We propose that participatory design involving teachable machines facilitates problem formulation activities. To test this, we integrated problem reduction heuristics into storyboarding and invited a university-based intergenerational design team of 10 children (ages 8-13) and 9 adults to co-design a teachable machine. We find that children draw from personal experiences when formulating AI problems; they assume voice and video capabilities, explore diverse machine learning approaches, and plan for error handling. Their ideas promote human involvement in AI, though some are drawn to more autonomous systems. Their designs prioritize values like capability, logic, helpfulness, responsibility, and obedience, and a preference for a comfortable life, family security, inner harmony, and excitement as end-states. We conclude by discussing how these results can inform the design of future participatory AI activities. 

\end{abstract}



\begin{teaserfigure}
\centering
\includegraphics[width=\textwidth]{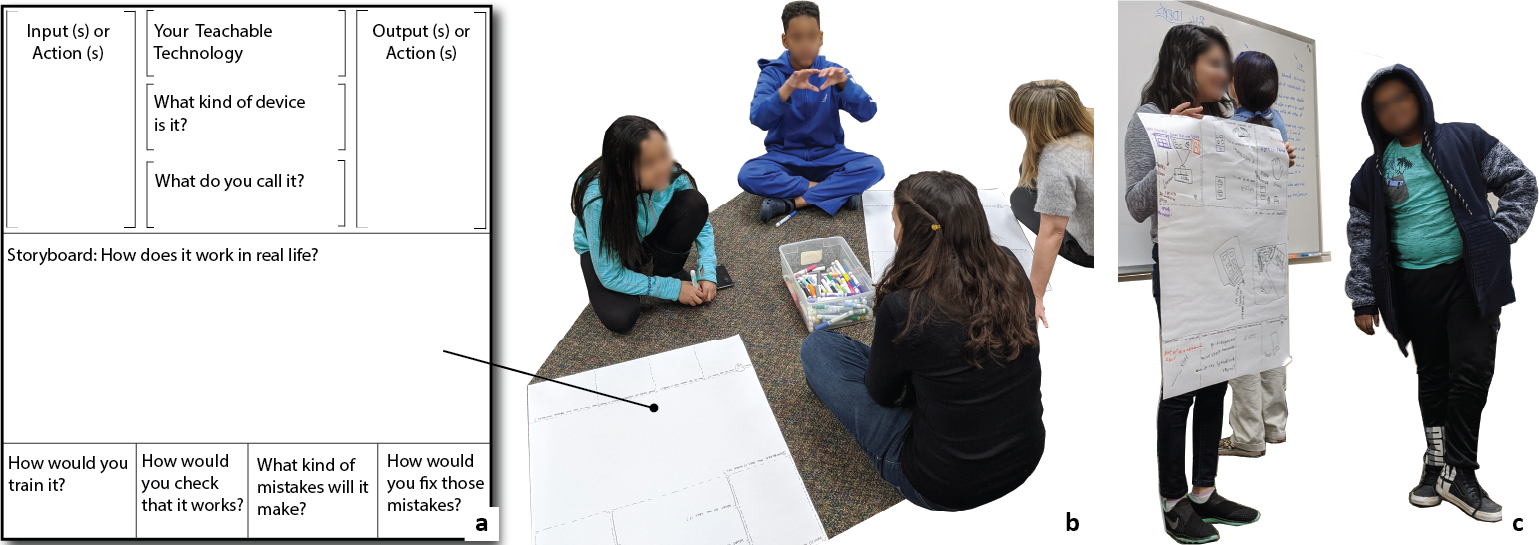}
\captionof{figure}{Our co-design study comprises: a) modified \textit{Big Paper} with a structured storyboard that b) child-adult pairs use to frame and discuss their ideas and then c) present their teachable machines while a researcher summarizes their input.}
 \label{fig:teaser}
\end{teaserfigure}

\maketitle

\section{Introduction}
Many recognize that ``the formulation of a problem is often more essential than its solution’’~\cite{einstein1938evolution}. In artificial intelligence (AI), problem formulation is a crucial step for developing technologies that maximize potential benefits and mitigate potential risks. Typically, problem formulation in this domain \invertedComma{involves determining the strategic goals driving the interventions and translating those strategic goals into tractable machine learning problems}~\cite{martin2020participatory}. Several practical resources exist to aid developers in AI problem formulation (\eg,~\cite{google2022formulate, amazon2022formulating}). However, recent research suggests that UX practitioners, including designers and product managers, are seeking additional support during the early stages of ideation and problem formulation in order to preempt AI product failures~\cite{yildirim2023investigating}.  
Given the rise of generative AI, problem formulation skills are being foregrounded as foundational, enduring, and adaptable competencies that ``might prove to be as pivotal as learning programming languages was during the early days of computing’’~\cite{acar2023ai}. Thus, it is important that we center problem formulation in activities that aim to help children develop AI literacies~\cite{dignazio2015approaches, druga20214as}. Yet,  to date there are few empirical studies of what these activities look like and how they can be structured for meaningful engagement.

In this paper, \textit{\textbf{we explore the affordances of participatory design activities with teachable machines\footnote{A term first coined by Andreae and Andreae~\cite{andreae1978teachable} in 1978 and resurfaced by Google Teachable Machine~\cite{google2017teachable} in 2017.} for engaging children in AI problem formulation. }}
Specifically, we hypothesize that teachable machines, where ``the user is a willing participant in the adaptation process and actively provides feedback to the machine to guide its learning''~\cite{patel1998teachable}, can play an essential role in facilitating such activities. We already see recommendations for incorporating teachable machines in formal educational activities that aim to spark curiosity and promote children’s understanding of how AI works (\eg,~\cite{touretzky2019learningactivities}). Yet, in these activities children tend to have \textit{little control} over the problem formulation~\cite{dwivedi2021introducing, vartiainen2021machine, hitron2019can}; children tinker with the input or output of \textit{predefined machine learning model} (\eg, a classifier) in a \textit{predefined context}  (\eg, recognize objects~\cite{vartiainen2020learning, dwivedi2021introducing} or gestures~\cite{hitron2019can}). 
These approaches are critical first steps.
However, we believe that by expanding the level of control and creativity children can exercise with teachable machines, activities can go beyond enabling children to uncover AI black boxes or change their perception of AI abilities~\cite{druga2021how}-- their current promise. 

To illustrate our approach, we pair problem reduction heuristics~\cite{maccrimmon1976decision} with a structured ``Big Paper'' storyboarding activity~\cite{fails2012methods}. Our modified Big Paper is shown in Figure~\ref{fig:teaser}a. We then invite a university-based intergenerational design team of 10 children (aged 8-13 years) and 9 adult partners to co-design their own teachable machines for a problem of their choice, as shown in Figure~\ref{fig:teaser}b. 
This storyboarding activity is situated within a larger effort to involve children in participatory machine learning activities that enable them to practice and build upon AI literacies. 
Our partnership with children involves first focusing their attention on specific aspects of the machine teaching process (\eg, similar to Dwivedi~\etal~\cite{dwivedi2021introducing} and Vartiainen~\etal~\cite{vartiainen2021machine}), then gradually removing such scaffolding so that they are equipped to tackle the broader landscape of problem formulation.
Within the context of this study, we ask:
\begingroup
\addtolength\leftmargini{-0.2in}
\begin{quote}
\textbf{RQ1}: What are the key aspects characterizing children’s formulated AI problems?\\
\textbf{RQ2}: What are the most prevalent design metaphors for AI in children’s ideas?\\
\textbf{RQ3}: Which personal values do children incorporate into their designs?
\end{quote}
\endgroup

Our study, exploratory in nature, provides empirical evidence on the affordances of participatory design activities with teachable machines for engaging children, as young as 8 years old, in AI problem formulation.
We find that children's formulated AI problems draw from their life experiences, addressing needs at home like security, automation, and familial support, as well as challenges at school such as math, safety, and social interactions. They envision AI with voice capabilities and often expect constant video monitoring. While some stick to supervised classification, others explore various machine learning approaches including learning by demonstration or unsupervised learning. Most anticipate errors and plan to address them through methods like adding more training data, retraining or, in some cases, imagining self-debugging capabilities.

We also find that most of the children’s ideas encompass a recent call on human-centered AI for a shift in the design metaphors around AI development to move away from ``autonomous AI systems’’ towards systems that ``center human capabilities and involvement’’~\cite{chan2021supporting}. 
Yet, the concept of intelligent agents and social robots who operate as teammates or with assured autonomy~\cite{shneiderman2022human} remained enticing for a few of the children; this was especially the case for those who described their AI-infused technologies as something that they would not have to train. 

Last, we used the Rokeach Value Survey~\cite{rokeach1973nature} as an analytical framework to examine the values reflected in children's designs. 
The Rokeach framework includes terminal values, which represent the ultimate end goals people strive for, while instrumental values reflect the desirable modes of conduct that people exercise to reach those end goals.
We found that instrumental values such as ``capability,'' ``logic,'' ``helpfulness,'' and ``responsibility,'' were shared among all children as preferred modes of behavior for their machines. 
``Obedience'' was also common. 
Less frequent were ``self-control,'' ``courage,'' and ``cheerfulness.'' 
Terminal values were also less present with ``a comfortable life'' being most prevalent followed by ``family security,'' ``inner harmony,'' and ``an exciting life.''

Our intertwined, two-fold aim of co-design with children illuminates their conceptualization of, and interest in AI-infused systems and also promotes designers' efforts to enhance future AI designs.
Our work thus aims to make the following contributions: (1) a structured storyboarding activity for AI problem formulation; (2) empirical results on how children (as young as 8 years old) formulate AI problems; (3) new insights on prevalent AI design metaphors among children that engage with teachable machines; (4) empirical results on shared instrumental and terminal values reflected in children’s designs of AI-infused technology; and (5) novel analytical lenses in AI problem formulation based on Shneiderman's design metaphors~\cite{shneiderman2022human} and Rokeach's value survey~\cite{rokeach1973nature}.

\section{Related Work}
In this section, we survey recent efforts in participatory AI and AI literacy learning. 
We explore how teachable machines are a means for children to practice AI literacies and to participate meaningfully in AI problem formulation.We close by discussing values' frameworks and their utility as an analytical tool for revealing values reflected in children's designs.

\subsection{Problem Formulation via Participatory Machine Learning}
There is an increasing literature in participatory machine learning that engages adults, those who typically have domain knowledge but may lack AI expertise, in the problem formulation stages \eg,~\cite{birhane2022power, prabhakaran2020participatory, delgado2022legal, sendak2020real}. Fewer studies focus on children.
The modified ``Big Paper'' in our work is informed by one of them, Woodward \etal~\cite{woodward2018using}, who used storyboarding with children across multiple sessions and emphasized anticipating and responding to errors. A critical difference lie in the fact that children's designs were constrained to a predefined problem space (a touchscreen handwriting game) and discussions around errors centered on existing technologies (voice assistants). 
While none of Woodward \etal's~\cite{woodward2018using} sessions included teachable machines, their findings strengthen our belief in the unique opportunity for teachable machines to involve children in AI problem formulation. Specifically, they found that children think of interactive systems in terms of ``both user input and behavior and system output and behavior.'' The input/output structure in teachable machines aligns well with this mental model. 

A second example comes from Vartiainen \etal~\cite{vartiainen2021machine}. Their study, which ran in parallel to ours, also engaged children in machine learning problem formulation via teachable machines. Similar to our work, children interacted in multiple sessions with a teachable machine (Google Teachable Machine~\cite{google2022teachablev2}), then used a design template to imagine their own teachable applications. Small differences between our studies lie in children's ages (8-13 in ours vs. 12-13 years old in~\cite{vartiainen2021machine}) and group formation (child-adult pair in ours vs. teams of children in~\cite{vartiainen2021machine}). More critical differences that highlight the complementary nature of these studies relate to design constraints and analysis approaches. For example, children's designs in Vartiainen \etal~\cite{vartiainen2021machine} strictly conformed to a multi-class single-label classification problem with input being image, sound, or pose in a given form factor (a laptop) and interface (Google Teachable Machine); in our study there were no such constraints. In contrast to Vartiainen \etal~\cite{vartiainen2021machine}, we use a design-metaphor and human-values lens to analyze children's engagement in problem formulation. 

A more recent work from Druga~\etal~\cite{druga2022family} strengthens our hypothesis and study design. They invited parent-child pairs to complete a series of activities along four sessions. Activities corresponded to the following AI literacies dimensions: multimodal situated practice, embodied situated practice, AI conceptual learning, critical framing of AI, and design future meaningful use. Similar to our study, activities for \textit{situated practice} involved teachable machines. More so, they preceded activities corresponding to \textit{AI conceptual learning }(\eg, ``Draw What is Inside'') and \textit{design future meaningful use} (\eg, ``Design AI'') that best align with AI problem formulation. 
Even though these last activities were constrained in the context of smart assistants and the focus was on parents’ roles in helping their children develop AI literacies, the results indicate that they were engaging and effective.

\subsection{Engaging Children with AI via Teachable Machines}
Teachable machines are gaining traction in specific contexts such as accessibility, and aging, where users have domain knowledge (\eg access needs) but may lack expertise in AI. For instance, they enable users to personalize object~\cite{kacorri2017teachable, garcia2017handson}, sound~\cite{bragg2016personalizable, jain2022proto}, speech~\cite{google2021euphonia}), and activity~\cite{kim2022mymove} recognition models built in their smartphones or smartwatches by fine-tuning them with their own data. On the other hand, systems like Wekinator~\cite{fiebrink2009wekinator} and Google's Teachable Machine~\cite{google2017teachable} are
more open-ended and encourage anyone to explore and fine-tune a machine learning system to their own use-cases or creative applications like creating a musical instrument~\cite{fiebrink2009wekinator}.
Upon analyzing use cases of Google's Teachable Machine, Carney \textit{et al.} found that it empowered educators and hobbyists, those who don't have expertise in AI, by making it easier to get started with machine learning~\cite{carney2020teachable}.
Thus, it is not a surprise to see researchers envisioning opportunities for teachable machines to increase children's familiarity and creativity with AI. 
One of the earliest studies was that of Dwivedi \etal~\cite{dwivedi2021introducing}, where children between the ages of 7-13 years old, used Google's Teachable Machine~\cite{google2017teachable} to explore machine learning concepts. Many studies followed (see~\cite{dwivedi2021introducing} for a comprehensive analysis). Some of them \eg, Vartiainen \etal~\cite{vartiainen2020learning}, involved even younger children, aged 3-9 years old. 

Learning activities with teachable machines are also recommended by the K-12 guidelines for AI education~\cite{touretzky2019learningactivities}. Yet, we see few instances of teachable machines in repositories of formal learning activities designed for use in K-12 and undergraduate courses.
For instance, our analysis of the Model AI Assignments repository\footnote{The repository is populated annually for the past 12 years by a workshop that collects and structures various assignments to help instructors~\cite{neller2010model}.}~\cite{modelai2022aaai}, hosted by the Association for the Advancement of Artificial Intelligence, indicates that only 2 out of the 79 assignments incorporate teachable machines. 
This could be partially explained by the overall low number (9 out of 79) of assignments involving \textit{younger children} or \textit{non computer science majors}, where teachable machines could be most helpful.
The rest require learners to read or edit code in \eg, Python or Java. 
Many are skewed toward undergraduate students in computer science with a focus on concepts like metrics or path-finding algorithms~\cite{modelai2022astar}. 
While these higher-level programming activities promote learning goals for computational thinking, they do not support much young, non-programmers' (8-13 years old) efforts to explore AI problem formulation, the goal of our paper.

The \textit{Introducing AI} assignment~\cite{modelai2022introducingai}, one of the two that included teachable machines, is the most similar to our work. After being introduced to computational thinking and AI literacy concepts like ``algorithm'' and ``data,'' children 6-9 years old construct their own imaginary AI device and investigate its weaknesses.
Similar to our work, this activity empowers children to explore AI concepts and apply their knowledge to design their own applications. 
This activity was implemented with several constraints imposed by the researchers: a set of predefined algorithms, sensors, and data.
Moreover, children's creations were not analyzed in terms of reflected values or design metaphors.
In contrast, we do not constrain the children's ideas by system capabilities, instead we ask them to consider a problem first, identify the training data needed, and establish a test and error-correcting process.
In addition, we gain deeper insights into children's ideas using Rokeach's value survey~\cite{rokeach1973nature} and Shneiderman's design metaphors~\cite{shneiderman2022human} as an analytical tool.

\subsection{Children's Values in Design}
Technology is inherently value-laden~\cite{eriksson2022teaching, verbeek2011moralizing, iversen2012values-led, van_mechelen2014applying}. 
Computing research on values has largely focused on values of ethical import~\cite{muller1997toward,friedman1996value-sensitive} such as privacy and security, as well as values of a deontological or moral import~\cite{han2022aligning}. 
While these studies are critical, research on individual, material values remains sparse~\cite{voida2005conveying,ferrario2016values-first,mougouei2018operationalizing,han2022aligning}. 
Those that do focus on personal values largely emphasize the values of adults and researchers rather than children~\cite{skovbjerg2016being,yarosh2011examining,kawas2020another,van_mechelen2014applying}, let alone children who are from historically marginalized communities. 
Skovbjerg \etal~\cite{skovbjerg2016being} recommend that Child-Computer Interaction (CCI) researchers acknowledge and discuss the embedded values in systems. 
This is especially important for AI problem formulation.
Some of the most important implications of a system can emerge during this early stage and raise profoundly different ethical concerns, such as possible threats to fairness and civil rights ~\cite{passi2019problem}. 
Thus, it is critical to examine and explicitly foreground children’s personal values in the design of AI-infused technology, and technology more broadly~\cite{elsayed-ali2020designing,druin2002role,yarosh2011examining,skovbjerg2016being,spiel2018micro-ethics}. 
This is increasingly important as personal values play a significant role in the design process~\cite{skovbjerg2016being} and children’s value trajectories are changing with societal changes~\cite{yarosh2011examining}. This may also help us attain a clearer understanding of what it means to align AI with human values and ways to go about this.

Social psychologist Milton Rokeach defines values as enduring beliefs and personal standards that guide and determine actions, attitudes, ideologies, judgments, justifications, and presentations of self~\cite{rokeach1973nature}. In his established value survey (RVS), Rokeach identified 36 values: 18 instrumental values reflecting preferred behaviors and 18 terminal values reflecting preferred end-states of existence~\cite{rokeach1973nature,iversen2012values-led,voida2005conveying,elsayed-ali2020designing}. 
Despite its rigorous validation and importance in value research—having laid the foundation for other frameworks such as Hofstede’s Cultural Dimensions and Schwartz’s Value Survey (SVS)—the RVS has been seldom used in computing research: only in five papers across 40 years~\cite{voida2005conveying,he2010one,anderson1978value,kayal2018automatic,munson1979values} and once in CCI research~\cite{elsayed-ali2020designing}. 
He \etal~\cite{he2010one} used the RVS as an analytical tool in broader human-centric computing literature analyzing persuasive energy feedback technologies. 
Elsayed-Ali \etal~\cite{elsayed-ali2020designing} used the RVS in CCI literature to scaffold children's designs of novel technologies. 
Until now, however, the RVS has not been used as an analytical lens in CCI research; other value frameworks have. 
For example, one study examined human values in adopting ubiquitous technology to support attendance control service in a primary school using the SVS~\cite{isomursu2009examining,schwartz1992universals}. 
Values including Benevolence, Achievement, Power, Conformity, and Self-Direction were identified as the values exhibited by children’s behaviors in adopting the service~\cite{isomursu2009examining}. 
However, the SVS is more concerned with motivational goals~\cite{schwartz1992universals} and Hofstede's model is more concerned with cultural values within organizational settings~\cite{hofstede_cultures_2001}, as opposed to Rokeach's focus on individual values and guiding principles in life. 
In this study, we use the RVS as an analytical tool to examine the perceived values reflected in children's designs. 
We aim to uncover the values and interests children embed in their everyday AI systems, especially during the problem formulation phase. 
As values give expression to human needs~\cite{rokeach1973nature}, we must strive to understand the values children embed in the AI technologies they design to better meet their needs~\cite{elsayed-ali2020designing,skovbjerg2016being,eriksson2022teaching,van_mechelen2014applying}.

\section{Methods}
\begin{figure*}
    \centering
    \includegraphics[width=.65\textwidth]{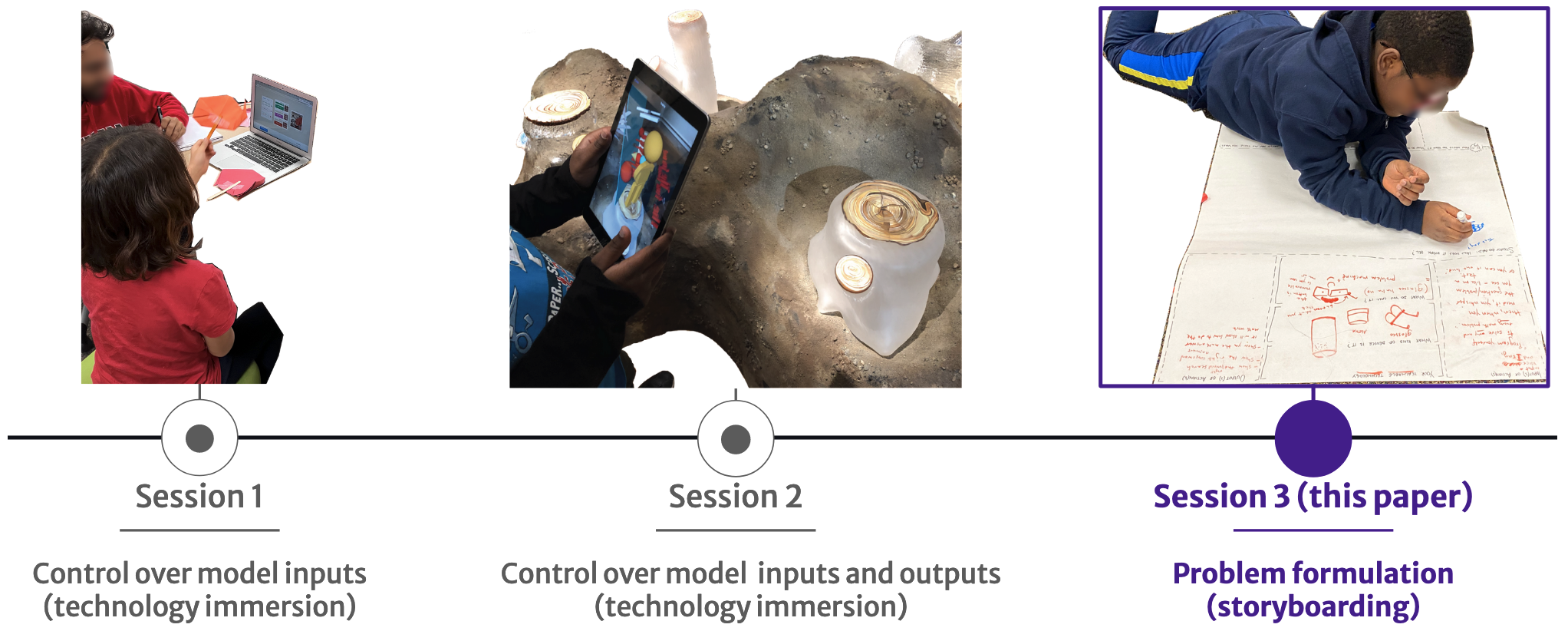}
    \caption{Our work is part of multiple co-design sessions. In session 1, technology immersion ~\cite{druin1999cooperative}, children engaged with Google’s Teachable Machines, trained the classifier by exerting control over the input (training examples of the objects to be recognized), and tested each other’s training efforts.
    In session 2, also a technology immersion,  children engaged with a teachable augmented reality application in a museum. 
    The application, developed by our team, enabled children to control both the input (training examples of objects in a museum exhibit) and the output (their own 3D designs that were triggered upon successful recognition of the object). 
    In this study, we present results from session 3, that engaged children in AI problem formulation via a modified ``Big Paper'' storyboarding activity. 
    In session 3, children explored how they might formulate and approach a machine teaching problem of their own design, influenced by their every day life experience.}
    \label{fig:sessions}
\end{figure*}

This work is part of a larger series of co-design sessions in which we explored children's conceptualizations of AI with machine teaching (see Figure ~\ref{fig:sessions} for an overview). 
In our first two sessions, we constrained the children's interactions and exposure to teachable machines.
Children tinkered with ML inputs in the first session (using Google's Teachable Machine~\cite{dwivedi2021introducing})
and created personally meaningful outputs in the second.
Specifically, for the second session, children visited 
a local museum exhibit where they trained their teachable machine within an augmented reality app, which would then display their own 3D artworks upon encountering specific artifacts from the larger art installation.
These experiences with similar machine learning models (3-way image classifiers) sensitized children to specific input/output aspects of the machine teaching process.
Building from this grounding in machine learning models, we then invited children to imagine how they might incorporate teachable machines and machine teaching use cases into their every day lives.
Our approach is aligned with established life-relevant learning approaches~\cite{clegg2012technology} that help children see the importance of scientific and AI literacy practices such as problem formulation in daily life.

\subsection{Child and Adult Co-designers}

\begin{table}[b]
  \small
    \centering

    \begin{tabular}{cccccc}
    \textbf{Pseudonym} & \textbf{Gender} & \textbf{Age} & \textbf{Race/Ethnicity}\\
 \toprule
    Brian & M & 8 & Black/African American\\ 
    \rowcolor{gray!15}
    Ed & M & 8 & Asian/White biracial\\
    Luke & M & 8 & Asian American\\ 
     \rowcolor{gray!15}
    Nancy & F & 8 & White/Caucasian\\ 
    Ollie & F & 8 & Asian American\\ 
     \rowcolor{gray!15}
    Adrian & M & 11 & Asian/White biracial\\ 
    Alan & M & 11 & Black/African American\\  
   \rowcolor{gray!15}
    Denny & F & 11 & Black/African American  \\ 
    Penny & F & 11 & Black/African American\\ 
    \rowcolor{gray!15}
    Kevin & M & 13 & Black/African American\\ \hline 
    \end{tabular}
    \caption{Pseudonym and demographics of the children.}
    \label{table:participants}
\end{table}
Our work is part of a multi-part project exploring machine teaching and AI literacy activities with children (see Figure \ref{fig:sessions}). 
In this study, 10 children aged 8-13 years old (4 girls, 6 boys) and 9 adult co-design partners (7 women, 2 men) participated.
The first two sessions (Figure \ref{fig:sessions}) afforded all the children an opportunity to gain hands-on familiarity with specific aspects of the machine teaching process; however, none of the children had coding experience beyond a handful of Hour of Code activities~\cite{code2023hour} at their schools.
All personally identifiable data has been removed to protect the children’s anonymity. 
Children's pseudonyms, gender, age, and race/ethnicity are shown in Table~\ref{table:participants}.
Overall, the children's demographics reflect our effort to include more non-dominant youth (i.e., non-White, immigrant backgrounds~\cite{mark2016psychology}), historically underrepresented in STEM learning, as co-design partners in the design of new technologies.
With regards to adult co-designers, two adults had a machine learning background (\eg, coursework and research experience) while others had backgrounds in human-computer interaction and education.
All adult co-designers have higher education qualifications.

\subsection{Selection and Participation of Children}
This study’s child participants are part of a university-based participatory design team. 
The children are recruited by word of mouth, from neighborhoods and municipalities local to the University of Maryland, based on family interest.
Children are selected from a wait list, which is always open to new and prospective child members, with a goal of balancing the children’s age ranges (typically 7-12 years old) and gender. 
In the case of the 13-year old for this particular study, the child started the academic year as a 12-year old and turned 13 during his year of co-designing with the team.
The design team obtains parental consent and child assent at the start of each academic year, and all child participants are protected under university Institutional Review Board approval (\#357390).
We obtained signed parental consent and child assent, including audio/video recording consent.
The adult researchers review participatory design and study goals with parents and remind the child participants that they can stop participation at any time during the co-design sessions. 
All personally identifiable data was removed to protect the children's anonymity.

\subsection{Study Design \& Rationale}
All co-design sessions across our study followed a Cooperative Inquiry-based~\cite{druin1999cooperative} approach to understand children's experiences with machine teaching and problem formulation, and each adhered to a similar 3-act structure~\cite{guha_cooperative_2013}:

\textit{Circle Time:} A warm-up to establish context and guide discussion. 

\textit{Main Design Activity}: The larger team forms smaller groups of child-adult co-designers (typically in adult-child pairs or 2-3 children per adult) in a focused design activity that aims to uncover how children imagine emerging technologies or learn, perceive, appropriate, and evaluate existing designs.

\textit{``Big Ideas''}: Each group shares their design ideas in a whole group presentation to surface, summarize, and synthesize common themes and potential design requirements~\cite{guha_cooperative_2013,fails_methods_2012}.

\textbf{Circle Time}: Specifically for the storyboarding session, all co-designers (children and adults) answered the question, \textit{Imagine you are helping a friend learn to do something. As they are learning, how would you help them fix their mistakes?}
This prompt enabled the team to consider challenges and successes related to how people make mistakes and how we can help fix them. For example, Alan shared that he would \textit{``probably tell them what's wrong and what's right''} and Penny said that she would \textit{``explain what they did wrong.''}
The team also discussed the potential for a variety of everyday applications of machine teaching by exploring examples from prior image recognition work~\cite{karmann2017objectifier}.

\textbf{Main Design Activity}: 
The main co-design activity consisted primarily of a modified ``Big Paper'' technique. 
We chose to employ storyboarding over other co-design techniques (such as prototyping, sketches only, or brainstorming with post-it notes) because our goal was to support children's freedom to brainstorm in problem formulation, and storyboarding is a graphic visualization that can depict imagined user scenarios or sequences early in the design process~\cite{fails2012methods, guha2013cooperative}.

Typically, co-designers use large, blank sheets of paper for their ``Big Paper'' storyboarding. 
Often, however, more structured variations of storyboarding have been found to support younger children's ideation process and design volume~\cite{fails2012methods}. 
For example, Comic-boarding uses a \invertedComma{familiar construct, the comic} to to capture their ideas within comic strip panels, which has been found to increase the volume of children's ideas during brainstorming~\cite{moravejiComicboarding2007}.

Similarly, to scaffold and surface the children's ideas regarding potentially complex problem formulation, we structured the ``Big Paper'' technique to employ problem reduction heuristics~\cite{volkema1983problem} by including sections that listed key questions as explicit design considerations (see Figure \ref{fig:teaser}a). 
Specifically, we included sections for (1) breaking down the machine learning process into functional units (\eg, inputs and outputs); (2) specifying the technology used; (3) imagining how it works in real life; (4) explaining how to train it; (5) anticipating how and when their machines might make a ``mistake''; and, (6) ensuring their systems could recover from mistakes.

Children constructed their storyboards in collaboration with adult co-designers; each child worked with an adult co-designer. In this study, both the children and adults took on the roles of \textit{design partners}, meaning they were both equal stakeholders and active participants throughout the design of the teachable machines~\cite{yip2017examining,bonsignore2013embedding}. 
Additionally, there were multiple advantages to pairing children with adult co-designers. 
First, this pairing supported a more balanced partnership through elaboration~\cite{yip2017examining,bonsignore2013embedding}, meaning children and adults worked collaboratively to generate and mix ideas together.
Prior CCI research with children has found retrospectively that children appreciate and benefit from the complementary support roles that adults take on in co-design~\cite{yip2017examining} to make space for and elaborate upon children's ideas~\cite{mcnally2016cciethics} and that the resulting design ideas may be more inclusive, open-ended, and nuanced than expected with complex topics such as privacy~\cite{kumar2018codesignprivacy}.
Second, adults provided children individualized attention and help with time management, structuring questions and childrens' responses via storyboarding. 
Third, the adult-child pairing promoted children’s critical reflection. Adults encouraged children to engage in reflection-in-action~\cite{schonReflectivePractitionerHow1983} by asking them probing questions like \textit{How would you test that it works?} and \textit{How does it fix its mistakes?}. 
Questions like \textit{Would violence or stealing be nice or good things to do?}, afforded the group with opportunities to discuss and reflect upon potential ethical shortcomings in the childrens' designs.

\textbf{Big Ideas}: During their ``Big Ideas'' presentation, (see Figure~\ref{fig:teaser}c), children elaborated upon their overarching idea and explained how to train, test, and tackle problems that they anticipated. 
One adult captured design ideas from each adult-child dyad on a whiteboard for everyone to see~\cite{fails_methods_2012,guha2013cooperative}.
For most co-design approaches with children, particularly with the cooperative inquiry method, the smaller sub-groups assemble together again after the main co-design activity to share and compare their respective creations (e.g., sketches, stories, or prototypes). 
This \invertedComma{Big Ideas} process can uncover emerging common themes, overlapping ideas, unique perspectives, help developers prioritize or rank requirements~\cite{fails_methods_2012}, and enable both children and adults to engage in critical dialogue about their designs~\cite{guha_cooperative_2013}. 
For example, during \invertedComma{Big Ideas}, many teams discuss what is technically possible, what design metaphors might underlie a seemingly whimsical idea, or what might be feasible alternatives and design requirements~\cite{guha2013cooperative}.

\subsection{Data Collection and Analysis}
We used multiple cameras to record the sessions and two researchers kept field notes.
Overall, our data consisted of 10 \invertedComma{Big Paper} storyboards, more than 3.5 hours of video recordings, more than 1.5 hours of audio recordings and 3 sets of field notes. 
Via a thematic analysis~\cite{clarke2015thematic} we aim to (RQ1) understand how children conceptualize, experience, and reflect on their engagement with AI problem formulation via teachable machines; (RQ2) uncover design metaphors for AI that are most prevalent in children's ideas; and, (RQ3) examine and explicitly foreground children’s personal values in their AI designs. We selected verbatim quotes from children to support our findings. Below, we outline our steps.

    \textbf{Problem Formulation}: Our structured storyboard was a silhouette of the design decisions that the children made, with a focus on the machine teaching process. 
    The structured activity employs a series of decision strategies~\cite{volkema1983problem} to reduce complexity in the problem formulation phase such as factoring into subproblems (input and outputs); determining problem boundaries (device, use cases, training, testing); and examining changes (anticipating and fixing errors). 
    One researcher used this lens initially to code problem formulation themes in the data; then a second researcher, in discussion with the first, iteratively refined the themes and sub-themes.

  \textbf{Design Metaphors}: Shneiderman's \cite{shneiderman2022human} recent overview of emerging theories on human-centered AI highlights systems' design approaches that ``\textit{support human self-efficacy, promote creativity, clarify responsibility, and facilitate social participation}''~\cite{shneiderman2020human}. 
  We situate our co-design efforts with children within this shift in perspective.
  Shneiderman describes how distinct design goals (science and innovation) \textit{``have value in thinking about design metaphors for future technologies''}~\cite{shneiderman2022human}, with science goals favoring ``\textit{more automation}'' and innovation goals favoring ``\textit{greater human control}.''  
  To explore how children relate to their machines, we mapped children's ideas into Shneiderman's four pairs of design metaphors: \textit{Intelligent Agents} vs. \textit{Supertools}, \textit{Teammates} vs. \textit{Tele-bots}, \textit{Assured Autonomy} vs. \textit{Control Centers}, and \textit{Social Robots} vs. \textit{Active Appliances}~\cite{shneiderman2022human}. 
  Similar to our iterative approach to unpack how children approached problem formulation, one researcher initially coded the data for evidence across the spectrum of \textit{``more automation''} in contrast with \textit{``human control''} and then iteratively refined the themes in discussion with a second researcher.

    \textbf{Human Values}: 
    We used the RVS~\cite{rokeach1973nature} as an analytical framework to examine the values reflected in children's designs. We developed an understanding of these values from past work by He \etal~\cite{he2010one} who used the RVS to classify persuasive energy feedback technologies. Similarly, we sought to employ the RVS to analyze children's values in AI problem formulation. Using Rokeach's 36 values (both instrumental and terminal) and their definitions, two researchers found patterns in the data that supported a smaller subset of these values. They individually coded the data, discussed classifications, resolved disagreements, and reached consensus.

\section{Results}
\begin{figure*}
  \centering
\includegraphics[width=\textwidth]{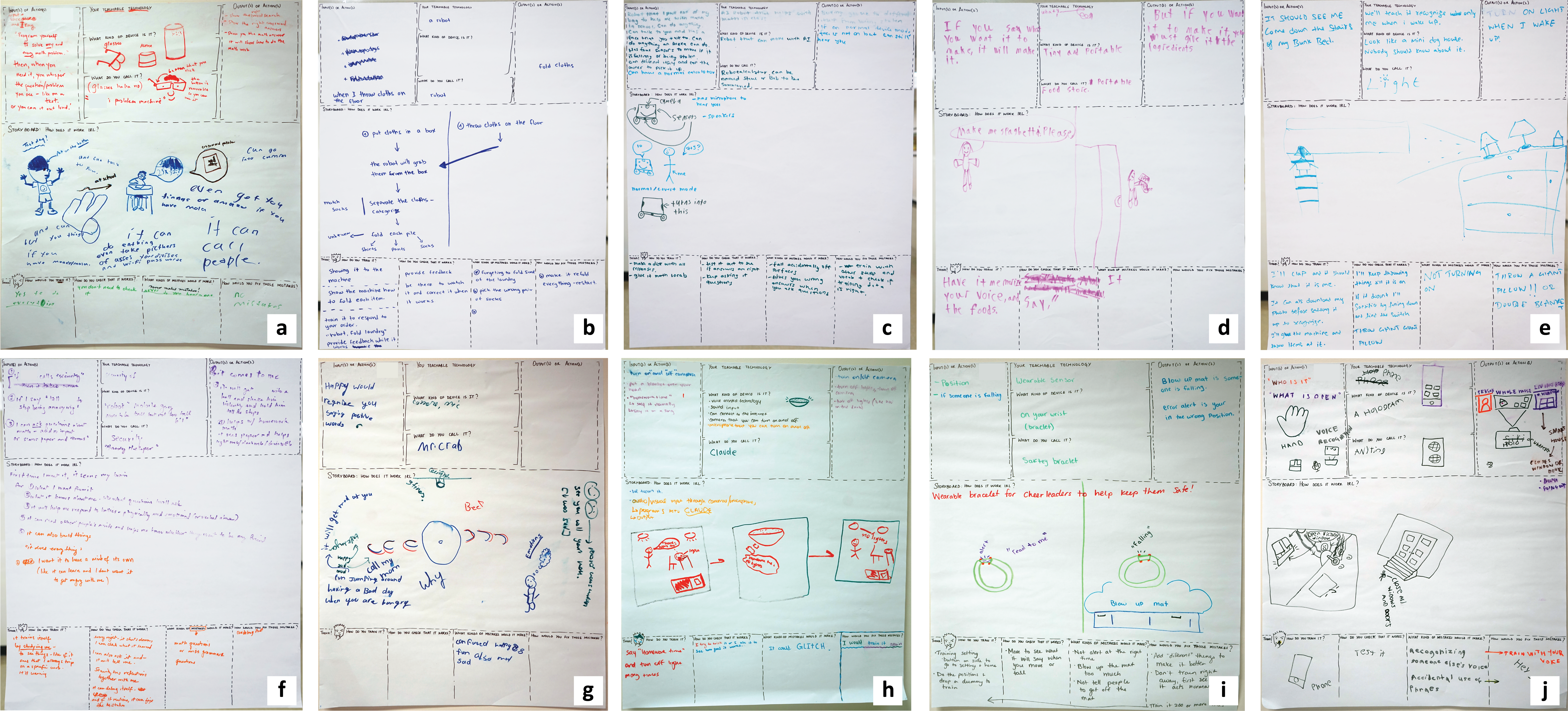}
 \caption{Children's storyboards: a) Brian's smart-glasses based ``\textit{Problem Machine}'' helps solve any math problem just by looking at it, b) Ed's robot helps you with chores such as folding clothes, c) Luke's \lineQuote{Robocalculator} can travel to you and solve math problems,  d) Nancy's portable food store makes food for you if you get the ingredients and tell it the recipe,
 e) Ollie's smart light controller, \lineQuote{Light}, recognizes her and turns on a light when she wakes up,  f) Adrian's robot security guard, ``\textit{Handy Helper},'' keeps his younger brother in check and protects him from bullies, g) Alan's wall-mounted emotion detection machine, ``\textit{Mr. Crab},'' responds to your emotions, h) Denny's smart light controller, ``\textit{Claude},'' creates your ideal homework environment,  i) Penny's safety bracelet for cheerleaders alerts you if you are about to fall and blows up a mat to cushion any fall, and j) Kevin's holographic smart home security system, ``\textit{Holo},'' alerts you of possible intruders and open windows/doors.}
 \label{fig:bigpaper}
\end{figure*}

To answer RQ1, we focus on the problem formulation aspect of children's designs, mainly how they describe purpose of their teachable machines, context, machine learning model, inputs and outputs, anticipated errors, and recovery from errors.
For RQ2, we then interpret children's ideas from the point of view of AI design metaphors. Last, to answer RQ3, we examine the designs and explicitly foreground the conversations for the human values they appeal to.
We present the conversations between adults and children, inline quotes from children, and the corresponding storyboards (see Figure~\ref{fig:bigpaper}) to support our findings.
The block quotes specify child or adult, where children's physical or non-verbal responses are within curly brackets and italicized, \eg, \{\textit{she nods}\} or \eg, \{\textit{points to the output section on the storyboard}\}. Any addition to complete missing words within the rest of the quote is denoted with square brackets (\eg, [the]). Apostrophes are used for inline quotes, which are also italicized. We highlight any part of a quote in \textbf{bold.}

\subsection{Children's Problem Formulation}

Children designed the following storyboards, as presented and ordered in Figure \ref{fig:bigpaper}, indicating the context that they would operate in as well as their input (for recognition) and output (for response). 
\begin{enumerate}[a)]
  \item \textbf{Brian's smart glasses-based Problem Machine} is a camera mounted on smart glasses that solves any math or crossword puzzle; ``\textit{you whisper it onto the glasses and it tells you or pops up [the answer] on your glasses.}''
  
  \item \textbf{Ed's robot that folds clothes} helps Ed to do house chores like folding clothes that are on the floor based on the type of garment like shirts, pants, and socks. It is called ``\textit{robot}''. If it makes a mistake Ed will ``\textit{make it refold everything.}''
  
  \item \textbf{Luke's ``\textit{Robocalculator}'' for solving math problems} can move around, it will come to you if you name it and call for it. The Robocalculator can be snuck in your schoolbag and you can ``\textit{tell it to whisper}'' answers to math problems.

  \item \textbf{Nancy's portable food store} can ingest ingredients and recipes for so you can order it to make food with commands like ``\textit{Make me spaghetti please.}''
  
  \item \textbf{Ollie's smart light controller called ``\textit{Light}''} can respond to claps close or open far away lights to help you avoid bumping into things at night. But, Ollie wanted ``\textit{it to make it difficult for other people}'' to use it. It only works when she claps, which ``\textit{Light}'' confirms by recognizing her face.
  
  \item \textbf{Adrian's robot security guard called ``\textit{Handy Helper}''} is a personal security guard that can bellow abuses to those who bully Adrian and even disciple his younger brother if he's being annoying. \textit{``Handy Helper''} is constantly monitoring Adrian and ``\textit{it can train itself by studying}'' Adrian. It also can help with math and homework.
  
  \item \textbf{Alan's wall-mounted emotion detector, ``\textit{Mr. Crab},''} can learn your emotions. It uses a camera and mic as inputs and can make video calls. If you are sad, it ``\textit{calls [your] mom}''.
  
  \item \textbf{Denny's smart light controller called ``\textit{Claude}''} controls the lights either by recognizing your activities (\eg, lights ON when you do homework) or by responding to your voice commands (\eg, ``\textit{homework time}'').
  
  \item \textbf{Penny's safety bracelet} alerts cheerleaders before they fall and inflates a mattress to cushion their fall. It also keeps track of who is on the mat and \textit{``tells people to get off the mat so that the faller can fall on the mat.''}

  \item \textbf{Kevin's home-security hologram called \textit{``Holo''}} is a full home-automation that responds to voice commands, acts upon alert sounds like fire alarms, and displays camera feeds of \textit{``who’s outside''}. For example, it recognizes a specific user's voice and closes or opens doors and windows in the house. 
\end{enumerate}

\subsubsection{Defining Model, Inputs, and Outputs}
In teachable machines, which fall within the broader machine teaching paradigm~\cite{zhu2018overview}, problems are often formulated as supervised machine learning tasks; specifically,  multi-class single-label classification. 
Thus, the teaching signal (\ie, input) typically refers to a labeled dataset of examples that the teacher (\eg, a child) provides the learner (\eg, a classifier). 
Similar to Dwivedi~\etal~\cite{dwivedi2021introducing} and Vartiainen~\etal~\cite{vartiainen2021machine}, the children in this study had previously been exposed to teachable machines that employed a single supervised classification model that mapped one type of input (\eg, images) to an output (\eg, GIFs or sounds).
Yet, many of the children's designs deviate from this machine learning example in their problem formulation. 
Even when children's designs employed supervised classification, each of their designs often included multiple such models capturing different functionalities.
For example, Penny's machine is capable of both (i) early fall detection (via the sensor in the bracelet) and (ii) collision detection (via weight sensors in the mat detecting presence of others).

Still, the concept of supervised learning appears even if the problem is not formulated as classification. Some children suggested that they would teach their teachable machines by ``\textit{showing it}'' the correct way to do a task, a common AI problem called \textit{learning by demonstration}~\cite{argall2009survey}. Specifically, these children's designs match problems formulated as \textit{behavior cloning}~\cite{goecks2020integrating}, where given limited data over a short time an agent can generate similar behavior.
Ed's robot would fold the clothes, and he would teach it how to fold a shirt or pants differently from folding a sock.
\begingroup
\addtolength\leftmargini{-0.2in}
\begin{quote}
\textbf{Adult}: How do you train the machine to do all of this?\\
\textbf{Ed}: By showing it.\\
\textbf{Adult}: By showing it. So what do you need to show it?\\
\textbf{Ed}: How to fold the clothes.\\
\textbf{Adult}: Do you need to show it, would you show it like how to fold a shirt or would you show it how to fold a sock? What would you show it?\\
\textbf{Ed}: How to fold all of them.
\end{quote}
\endgroup

The promise of teachable machines is that end-users can fine-tune them by training them. However, children often assume existing capabilities in their machines without having to explicitly train them. This could be either because others trained them (pre-trained) or simply because they are ``\textit{smart}''; surprisingly, these assumptions don't fall far away from recent discourse on foundation models even though the study was completed prior to the rise of ChatGPT~\cite{openai2022chatgpt} in the public discourse. 
All, except Penny and Ollie, assumed that their system comes equipped with built-in voice recognition, which may be related to their awareness of so-called intelligent agents~\cite{beneteau2020parenting} like ``\textit{Siri}'' and ``Alexa.'' 
Specifically, Kevin's Holo could automatically detect people without Kevin's instruction, Adrian's security robot would ``\textit{know}'' if he's being bullied, Nancy's portable food robot did not need a recipe just a command because it ``\textit{would know how to make it}'', and Brian's problem machine could solve any math or crossword puzzle without explicit training from him (and this was before ChatGPT's~\cite{openai2022chatgpt} rise in popularity.) 

We were also surprised to see a high variation among input types, with video being the most common.  
More importantly, children often envisioned a state of constant data collection/monitoring, providing a ripe opportunity for discussions around security and privacy in AI. For example, Adrian's Handy Helper will always be monitoring Adrian and learning (unsupervised) from him and Kevin's Holo will monitor his house with cameras,
\begingroup
\addtolength\leftmargini{-0.2in}
\begin{quote}
\textbf{Adult}: So it works with cameras in your house too?\\
\textbf{Kevin}: Yeah. Cameras like in the kitchen and living room.\\
\textbf{Adult}: So cameras throughout the house that it talks to.\\
\textbf{Kevin}: Yeah.
\end{quote}
\endgroup

\subsubsection{Conceptualizing Teaching Strategies}
For the machine capabilities that children decided to provide training examples, their teaching strategies diverged. Penny had a clear approach from the start. Each cheerleader would record their correct motion (\lineQuote{like 200 or more}) by using a button on the bracelet,
\begingroup
\addtolength\leftmargini{-0.2in}
\begin{quote}
    \textbf{Adult}: How do you train it to know the proper positions?\\
    \textbf{Penny}: You do the proper positions and like \{\textit{Penny places her right hand on her left wrist as if she is touching a button on the safety bracelet}\} you tell it to train.

\end{quote}
\endgroup
Apart from correct motions, Penny suggested in her storyboard that \lineQuote{you could drop a dummy to train it for falls}, closely related to the \textit{synthetic data} approach in machine learning often deployed when there are ethical concerns from using real data (\eg,~\cite{kim2022how}). 

The others vacillated on their machine's specific functions as well as what and how they needed to teach it. This is not a surprise, as problem formulation is often an iterative process.
We observed that during this process, children would typically add to the teaching signal to provide more \textit{context} or \textit{discriminatory features}. For example, Alan initially thought that Mr. Crab would recognize (\lineQuote{hunger}, \lineQuote{happy}, \lineQuote{fun}, \lineQuote{sad}, and \lineQuote{mad}) based on his voice and his training labels like \lineQuote{I am hungry}. Then, he added that physical signals like \lineQuote{jumping} could also be indicators: 
\begingroup
\addtolength\leftmargini{-0.2in}
\begin{quote}
    \textbf{Alan}: Yeah and just tell when you're hungry by saying \lineQuote{I'm hungry!} You say it out loud ... And it would order you ... anything you want, or get you something from your fridge.
\end{quote}
\endgroup
At this point, he adds \lineQuote{saying positive words} to the storyboard and proceeds to explain how he would teach it,
\begingroup
\addtolength\leftmargini{-0.2in}
\begin{quote}
    \textbf{Alan}: It can tell when you're happy and having a good day or when you're jumping and you're excited or something like that...it recognizes by you saying positive words. You have to say it 6 times before it recognizes.
\end{quote}
\endgroup

While Penny and Ollie wanted a button or an app to label start and stop for recording the training moves or actions, other children used speech for labeling. For example, Alan prompted \lineQuote{I'm hungry now}, Denny wanted to say \lineQuote{Homework time!} and to start doing homework on the table, and Kevin wanted, \lineQuote{you go to each door and like tell it what door it is.}
We suspect children's labeling approaches are reflective of their interactions with existing voice technologies.

\subsubsection{Anticipating Failure}
All children presumed that there would be some pre-training done for their system which they could train upon with new data or scenarios (\eg, fine-tuning).
Hence the most common error they foresaw would be the likely scenario of insufficient pre-training.
For example, Alan anticipated that Mr. Crab would not be \lineQuote{properly trained}; thus, it would not be able to differentiate between \textit{similar} emotions like \lineQuote{happy} and \lineQuote{fun},
\begingroup
\addtolength\leftmargini{-0.2in}
\begin{quote}
    \textbf{Adult}: So what mistakes would it make? Do you think it will confuse sad with happy or?\\
    \textbf{Alan}: Probably happy and fun. They're kind of the same.\\
    \textbf{Adult}: What else do you think would be challenging?\\
    \textbf{Alan}:  Mad and sad.
\end{quote}
\endgroup

However, we observed that children anticipated errors in their designs that go beyond the components that they taught.
For example, when Penny discussed testing her safety bracelet with acrobatic moves, she anticipated that it might not alert at the right time, a \textit{latency} challenge, in contrast to classification challenges such as blowing up the mat too often (\textit{false positives}) and not telling people to get off the mat (\textit{false negatives}). 

Those who used voice to interact with their system (most of the children) brought up pairs of phrases with potential for confusion like \lineQuote{Ok cool} instead of \lineQuote{Ok Google} similar to existing listing of misrecognized wake words~\cite{fast2020tired}. Indeed, one child, Kevin, explicitly demonstrated a wake word mistake by pulling out his phone and saying a wrong phrase.
Kevin's Holo, supporting personalized phrases and commands by the home owner, could not only misrecognize some words but also mistake the identity of the person who uttered them,
\begingroup
\addtolength\leftmargini{-0.2in}
\begin{quote}
    \textbf{Kevin}: Some mistakes that it can make is if it recognizes someone else's voice, and to prevent that you would have to train it with your voice. And if you accidentally use a phrase like \lineQuote{Alarm} and let's say your friend said \lineQuote{My alarm just randomly turned on} and after that it would start an alarm randomly calling 9-1-1. You don't want that to happen! You would have to say like \lineQuote{Hey whatever the name is, do whatever.}
\end{quote}
\endgroup 

\subsubsection{Proposing Recovery}

While children's storyboards diverged among design ideas, teaching strategies and anticipated errors, they tend to converge when fixing errors: the majority (5) of those who opted to train their machines also opted to add more examples for any misrecognized scenario.
For instance, in the case of a misrecognized home owner, Kevin suggested adding more examples from that person's voice much like the initial training that is done by existing systems (\eg, saying \lineQuote{Hey Google!} many times).

Penny also said that she would add more examples but they would be \textit{``different''} ones hinting at variation in training,
\begingroup
\addtolength\leftmargini{-0.2in}
\begin{quote}
    \textbf{Adult}: So would you just keep training it [and] training it?
    
    \textbf{Penny}: No. Maybe we could add different things. 
\end{quote}
\endgroup

Perhaps it is not surprising to see this as the most common strategy \eg, many children have used it with Google Teachable Machine~\cite{google2017teachable}. 
Even adult non-experts opt for \textit{more data} as a fix~\cite{yang2018grounding}.

Out of all the children, only Denny and Luke would help their machines recover from errors by starting the training from scratch; in contrast, Brian expected his problem machine to make no mistakes at all as \lineQuote{it knows everything,}
\begingroup
\addtolength\leftmargini{-0.2in}
\begin{quote}
    \textbf{Adult}: How would you check if the answer [to the math problem] was correct?\\
    \textbf{Brian}: You don't need to check. There's a built in calculator, just in case you need to check.\\
    \textbf{Adult}: You don't think a calculator does not make a mistake? What if someone plugged in the wrong number?\\
    \textbf{Brian}: It's got a built in calculator, just in case. It never makes a mistake, the glasses never make a mistake \{\textit{for emphasis, Brian underlined} \lineQuote{never} \textit{on his storyboard}\}.
\end{quote}
\endgroup
Like Brian, Luke and Adrian wanted their machines to solve math problems and never make mistakes.
When we concluded this study in October 2022, we had yet to anticipate that ChatGPT would exhibit this behavior of ``solving'' math problems when released in November 2022.
Children might be too impressed by ChatGPT, ``never failing'' to solve Grade 2-7 math problems, which could overshadow the limitations of this and other AI-infused technologies~\cite{cheng2023analyzing, rane2023enhancing, wardat2023chatgpt}.

In this section, we illustrated how children can construct different scenarios where a machine learning system can fail.
The following section discusses the design metaphors best describe children's AI-infused technologies and whether they retain oversight over their error-prone or never-failing machines.

\subsection{Design Metaphors Reflected in Children's Designs}

\subsubsection{\lineQuote{Intelligent Agents} and \lineQuote{Supertools}}
    We consider the first pair of Shneiderman's design metaphors~\cite{shneiderman2022human}: \textit{``Intelligent Agents''} and \textit{``Supertools''} (or \textit{``AI-infused Tools''}~\cite{chan2021supporting}) and how they relate to children's ideas.
    \textit{``Intelligent Agents''} are capable of thinking and making decisions with little or no oversight while
    \textit{``Supertools''} offers a high degree of human control and automate the more repetitive aspects of a task.
    The two are often thought of in contrast as there is a trade-off between more automation and greater human control. 
    Surprisingly, we did not see such contrast in many children's designs; 7 of 10 children's ideas had characteristics of \lineQuote{Supertools}. 
    Only Adrian had aspects of \lineQuote{Intelligent Agents} in his idea, while Brian and Alan had characteristics of both \lineQuote{Intelligent Agents} and \lineQuote{Supertools}, particularly the automatic solving of \textit{all} math problems and a \textit{human-level} detection of emotions.

During circle time, we showed children examples of people training a machine learning system to respond with a specific action (\eg, a person hiding under bed covers to trigger an image detector to turn off a light). They asked insightful questions; for example, Brian asked \lineQuote{where does the camera feed go?}
However, when discussing what their teachable machines \lineQuote{know} beforehand and what they are capable of, children's responses reflected a technology utopian perspective \cite{dickel2017logic, ghazizadeh2012extending}.
Their system had some \textit{magic,} or \textit{know it all} before any training was needed. 
For example, Brian noted that his machine would know all the math answers and not need training. Instead, it would train and help the human (\lineQuote{it's a computer! it already knows!}). 
Or, when Adrian, who's idea only had the aspects of \lineQuote{Intelligent Agent} described how ``Handy Helper'' would be independent to take its own decisions in a particular scenario based on human-like emotion recognition and mind reading capabilities, 
\begingroup
\addtolength\leftmargini{-0.2in}
\begin{quote}
    \textbf{Adrian}: It can tell sort of ... tell what I am thinking so then I can do ... really cool stuff. If somebody bullies me on the street and I’m like, ``I may be small BUT'' then Security can read my mind and become un-invisible and be like ``SURPRISE!'', slap.\\
    ...\\
    \textbf{Adult}: And then it will help me respond to bullying. Only physically or?\\
    \textbf{Adrian}: Physically and also I can give it emotional damage such as ``smackdown'' or something.\\
    \textbf{Adult}: So verbal abuse.\\
    \textbf{Adrian}: Verbal abuse, [because] words hurt more than weap-ons.\\
    \textbf{Adult}: So does it just reads ``a bully’s'' mind or other people you are talking to.\\
    \textbf{Adrian}: It can read any person’s mind. So if I meet anybody I can tell what they are thinking. 
\end{quote}
\endgroup
    
As \lineQuote{Supertools}, children's machines would control lights, call parents, inflate mats, and open windows, all on their own, by making appropriate inferences. All their machines (except Penny's bracelet) would also do these actions upon explicit user commands. The following quote from Kevin shows how he wanted \invertedComma{Holo} to automate opening doors and windows both on command and in response to a kitchen accident,
\begingroup
\addtolength\leftmargini{-0.2in}
\begin{quote}
    \textbf{Kevin}: Let's say you were cooking some eggs. And after that it just starts getting set on fire and smoke is going everywhere. And you want the window to be open. So all that bad smoke could go out. What you would do is that you would say \invertedComma{Open kitchen windows} or those carbon monoxide things would detect the smoke, like always, and it would start beeping and also it would open the window near it. Probably every window sometimes.
\end{quote}
\endgroup
Only Brian's problem machine had characteristics of both \lineQuote{Intelligent Agents} and \lineQuote{Supertools} because he had a superhuman expectation that it would never make any mistake on a math problem.
Independently of how the \invertedComma{intelligence} came to be, Brian's problem machine was there to help, a concept well aligned with the goals of Supertools to \invertedComma{amplify, augment, empower, and enhance humans}~\cite{shneiderman2022human}. Like Brian's idea, Vartiainen \etal~\cite{vartiainen2021machine} also noted that more than a few of the children's ideas related to homework automation.
They wanted a teachable machine that would \invertedComma{show mistakes or correct answers when taking a picture of homework.}

\subsubsection{\lineQuote{Teammates} or \lineQuote{Tele-bots}}
When looking at children's ideas under the lens of the \lineQuote{Teammates} and \lineQuote{Tele-bots} pair of design metaphors, we observed a shift towards the second goal. We found that only 1 child assumed human-human interactions in their designs, where machines would be perceived as teammates, partners, or collaborators. For example, all children leveraging voice interactions imagined them being one-sided commands rather than conversations, such as Denny's smart light controller that would respond to her command \lineQuote{Homework time!} but only Adrian wanted his \lineQuote{Handy Helper} robot to tell him if the person talking to Adrian wants to be his friend.
As he continues to describe how the robot can scan anyone's mind he says, 
\begingroup
\addtolength\leftmargini{-0.2in}
\begin{quote}
    \textbf{Adrian}: It can read any person's mind. So if I meet anybody I can tell what they are thinking. So if I am trying to make a friend. \\
    \textbf{Adult}: But what if?\\
    \textbf{Adrian}: So if they don’t want to be my friend then they just don't want to be my friend because I have a super awesome robot that is currently reading their mind.
\end{quote}
\endgroup
\lineQuote{Handy Helper} ensures that Adrian's younger brother does not annoy him. Much like a helping hand or a teammate, the \lineQuote{Handy Helper} warns him when ``\textit{it sees me about to make a mistake it can tell me ``Adrian don’t chase that dog around,}'' or like ``\textit{Adrian don’t grab that pointy stick.}'' These are all responses or alerts to a situation. Even when Adrian asks his robot to reflect on its mistakes and fix them, he does not imagine much of a conversation but just something it does automatically.

As Tele-bots, children's machines often provided \lineQuote{superhuman perceptual and motor support while allowing human–human teamwork to succeed}~\cite{shneiderman2022human}. For example, Kevin's Holo could signal ``\textit{who is outside}'' his home and Penny's bracelet would alert cheerleaders if they were about to fall to them and their teammates, 
\begingroup
\addtolength\leftmargini{-0.2in}
\begin{quote}
    \textbf{Penny}: For example you're holding someone or someone is in the air and you can't read it yourself you could tell it to read to you and it will tell you everything you are doing wrong, and it will also alert when someone is falling ...
\end{quote}
\endgroup
Another example is Ed's robot, that is assumed to have motor actuators and grasping capabilities that let it find and grab the clothes lying around, perceive what kind of clothes are folded in what way and then fold the clothes. 

\subsubsection{\lineQuote{Assured Autonomy} and \lineQuote{Control Centers}}
When looking at children's designs under the lens of the \lineQuote{Assured Autonomy} and \lineQuote{Control Centers} pair of design metaphors, we observe that children's ideas aligned more closely to the second goal. While systems with \lineQuote{Assured Autonomy} act independently, many children's machines instead depend on and support human control and oversight (supervised autonomy) through \lineQuote{Control Centers} or Control Panels~\cite{shneiderman2022human}.
Only 2 of 10 children had aspects of \lineQuote{Assured Autonomy} and and the rest had aspects of \lineQuote{Control Centers.} 
Children could predict the response of their machine learning system, maintain situation awareness, and take control of their machines at any time. For example, Kevin's Holo supported supervisory control and situation awareness,
\begingroup
\addtolength\leftmargini{-0.2in}
\begin{quote}
    \textbf{Kevin}: What it does is that if you wave your hand toward it, it pops up a hologram showing everything about your house. Over here, it would probably be like security cameras of the front door, even though there are more security cameras around the house, on the outside and the inside. It's just showing you the front door because that's the main place. There would also be a picture of the house—like [the] holographic picture of your house that you could rotate around.
\end{quote}
\endgroup

In contrast, Adrian gave complete discretion to his robot, and Nancy expected her food machine to make any food without supervision. Just put in the ingredients, and it does the rest.
Adrian's \lineQuote{Handy Helper} learn autonomously, ``it can train itself by studying me and seeing if I make any mistakes, like continuously.''
When the adult asked Adrian about the kind of mistakes his robot would make or how it could fix them, Adrian provided evidence of \lineQuote{Assured Autonomy} by highlighting an inherent issue with the concept of complete autonomy.

\begingroup
\addtolength\leftmargini{-0.2in}
\begin{quote}
    \textbf{Adrian}: ... I want it have a mind of it own. Right. So do you know rainbow friends?\\
    \textbf{Adult}: No.\\
    \textbf{Adrian}: Okay. Rainbow friends is this roblox game. I guess you could just look it up. But, um, it’s like rainbow friends have minds of their own. I want my robot to have a mind of its own so I don’t have to be like ``why do you learn to [do something]''. Because it might turn against me and I don’t like it. 
    And I don’t like a 6 foot 4 like super robot shooting rocket launchers and stuff to hit me.
\end{quote}
\endgroup

In illustrating the tension between control and autonomy, Shneiderman provides the critical reason: that ``operators [should] have a clear mental model of what will happen next,'' \ie, \textit{predictable behavior}.
Adrian could use the storyboard's structure, the questions on mistakes, and fixing errors to critique his idea and push back on any unpredictable behavior.

\subsubsection{\lineQuote{Social Robots} or \lineQuote{Active Appliances}}
When viewing children's designs under the lens of the \lineQuote{Social Robots} and \lineQuote{Active Appliances} pair of design metaphors, we find more evidence of the second design goal being prominent with some caveat. Children envisioned their machines as physical devices attached to a ceiling, placed somewhere as speakers, put on as glasses, worn as bracelets, or as holograms (\ie, of a house, not a humanoid), or humanoid like in Adrian's case. Many could listen and respond to commands. Alan, Kevin, and Denny named their machines. However, even those with human-like names like \lineQuote{Claude} and \lineQuote{Mr. Crab} did not have human-like forms. 
While having a human-like name and being able to listen/speak is at some level anthropomorphism~\cite{luger2016like}, these features are standard even in everyday devices and appliances that are not considered social robots (\eg our smartphones). 

Emotion, often associated with Social Robots, is prevalent in children's designs. For example, in Vartiainen \etal~\cite{vartiainen2021machine}, one of the apps would detect your mood using facial expressions and \lineQuote{if you are feeling sad, the app will comfort you.} Woodward \etal~\cite{woodward2018using} made a similar observation, that children wanted their intelligent user interfaces to recognize their emotions and respond with appropriate emotion, stating that \lineQuote{this type of intelligence was a fundamental part of many of the children’s designs}~\cite{woodward2018using}. 
We found that only two children, Alan and Adrian, have characteristics of social robots, particularly the ability to recognize emotions or read minds.
However, Alan's \lineQuote{Mr. Crab} or Adrian's \lineQuote{Handy Helper} could only recognize emotions rather than express them. When Alan needed comfort, he wanted it to come from his mother, not \lineQuote{Mr. Crab},

\begingroup
\addtolength\leftmargini{-0.2in}
\begin{quote}
    \textbf{Alan}: …and you can add contacts to it\\
    \textbf{Adult}: Oh cool\\
    \textbf{Alan}: So then it can call my mom
\end{quote}
\endgroup
When presenting his idea to the larger group, he said that \lineQuote{Mr. Crab} would call his mom when he's sad because, \lineQuote{when you're sad, I just call my mom, basically. I just call her.}
In contrast to Social Robots mentioned by Shneiderman, \lineQuote{Mr. Crab} plays the intermediary role and recognizes Alan's emotional state and then automatically calls his mom rather than engaging Alan in a conversation or having synthetic fur for him to touch.

The rest of the ideas, by 8 of 10 children, fall under \lineQuote{Active Appliances} with robotic characteristics like actuators in Ed's clothes folding robot or treaded robots like Luke's \lineQuote{Robo-calulator} would come to you when you called it.
The most interesting characteristic that made children's ideas an \lineQuote{Active Appliance} was how they would fix its mistakes.
As Shneiderman points out, there is ``much room for improvement in the frustrating designs [of Active Appliances], which are often internally inconsistent, difficult to learn, and vary greatly across devices.''
Much like our response to frustating appliances, when some people ``hit the TV'' or ``just restart it'' to fix it, Ollie would be annoyed if her \lineQuote{Light} didn't work causing her to stay stuck in the dark on her bed,
\begingroup
\addtolength\leftmargini{-0.2in}
\begin{quote}
    \textbf{Adult}: What will you do if it doesnt turn on?\\
    \textbf{Ollie}: Don’t worry even if it takes an hour to do, I'll still figure it out. I’ll be like here’s a pillow and keep on throwing things for an hour and sacrifice myself and go switch on the lights. And if it doesn’t work \textbf{I’ll get so mad I’ll scream into my pillow}.\\
    \textbf{Adult}: What do you think? What kinds of mistakes will it make?\\
    \textbf{Ollie}: Umm, probably not turning on the lights. That's the only mistake it can do. And maybe I can just, I have a giant giant pillow that I’ll throw at this so this turns on. Or just a giant blanket. Like a double blanket.
\end{quote}
\endgroup

\subsection{Values Reflected in Children's Designs}
When looking at the values represented in children's designs, we group our observations following Rokeach's framework~\cite{rokeach1973nature}, distinguishing between instrumental and terminal values.

\subsubsection{Instrumental Values in Children's Designs}
Rokeach describes instrumental values as preferred modes of behavior. For example, the value of ``capability'' entails competency and efficiency. Of the 18 instrumental values, the two coders agreed upon 10 instrumental values that were apparent in children's designs, with the following 4 values represented in all their designs: ``capability,'' ``logic,'' ``helpfulness,'' and ``responsibility.'' We also found the following values apparent in children's designs: ``obedience" in 6 of the designs (including all of the younger children's designs); ``cleanliness" in Kevin and Ed's designs; ``self-control" and ``courage" in Adrian's design; and ``cheerfulness" in Luke's design. Below, we provide examples of the 4 most frequent instrumental values in children's designs.

All of the children's designs appealed to the value ``capability,'' making explicit reference to how their system might increase their efficiency or competency, or instances where the system itself exhibits efficiency and competency. For example, Kevin's Holo would help him close the windows and doors of his house when he is going to sleep or outdoors, thus helping him complete the task more efficiently, 
\begingroup
\addtolength\leftmargini{-0.2in}
\begin{quote}
    \textbf{Kevin}: Let's say you're about to go to bed or go somewhere and you have no time to go around the house and close every single window because you live in a mansion. You could say ``What is open?'' Let's say there was a window all the way upstairs you didn't know about. What you would do is say ``What is open?'' and it would tell you every single door that is open and every single window that is open.
\end{quote}
\endgroup

All of the children's designs reflected the value ``logic,'' which Rokeach defines as being consistent and rational. Children primarily touched on how they would program the logic of their classifiers including the context, input, and outputs. For example, Penny would train her safety bracelet using different positions,
\begingroup
\addtolength\leftmargini{-0.2in}
\begin{quote}
    \textbf{Penny}: I can train it by going to the settings. It has a setting where you can go to...train it that way.\\
    \textbf{Adult}: How many times did you [want] to train it to make sure the mat blew up if you were the one falling?\\
    \textbf{Penny}: 200 or more, I'll train it 200 or more.
\end{quote}
\endgroup

All of the children's designs appealed to the value ``helpfulness,'' which means working for the welfare of others. Children described how their teachable machines would assist them in accomplishing a specific task, oftentimes explicitly using the word ``help'' to describe their machines. Adrian described his robot security guard as \textit{``a \textbf{handy helper}''. He’s a handy and he’s like a helper and he helps me with everything.''} Similarly, Penny stated \textit{``My machine is a wearable sensor, like a bracelet, and it’s a bracelet for cheerleaders to \textbf{help} them stay safe.''} 

Rokeach's value ``responsibility'' was reflected in all of the children's designs and means dependability and reliability. Children trusted their machines would work reliably and provide the intended outcomes. For example, Alan relied on his emotion detection machine to detect the correct emotion and respond to it accordingly. Luke also relied on his Robocalculator to not make mistakes, \textit{``I would make this special disk and this has everything. So when somebody ask[s] some question the disk will automatically go to its brain. And it's always right and never wrong.''} We see this value also in ability to recover from errors. For example, every night Adrians' robot ``\textit{will do his reflection}'' and could learn from mistakes.

\begingroup
\addtolength\leftmargini{-0.2in}
\begin{quote}
    \textbf{Adrian}: So I’ll explain to him patiently that it did wrong and it will be like “Oh okay.” And then we can, it can ask me to help it but if it doesn’t then it can figure it out himself. Then it can debug it.\\
    \textbf{Adult}: It can debug itself?\\
    \textbf{Adrian}: Yes. But if it needs help from me then I can help him.
\end{quote}
\endgroup

The value ``obedience,'' was also reflected in 6 of the children's designs, and entails dutifulness and respect as well as user control over the AI-infused system. Adrian used the analogy of a dog to describe how his robot security guard obeys his orders: \lineQuote{when I call ``Security'' it will come towards me. It’s like dogs when you call its name it will come to you.}

\subsubsection{Terminal Values in Children's Designs}
Rokeach describes terminal values as preferred end-states.
For example, the value ``family security'' entails taking care of loved ones. Of the 18 terminal values, the two coders agreed upon 4 present in children's designs. We classify children's designs as appealing to the following terminal values: ``family security,'' ``a comfortable life,'' ``inner harmony,'' and ``an exciting life.'' While values relating to family and comfort are of highest priority to children~\cite{elsayed-ali2020designing}, these values often go unacknowledged in existing AI discourse and frameworks. We call explicit attention to these values in our analysis. Below are examples of the 4 terminal values in children's designs.

The value ``family security'' was reflected in 3 designs belonging to the older children. Alan's machine called his mother when he felt sad; Penny's protected her teammates; and Kevin's safeguarded his home,

\begingroup
\addtolength\leftmargini{-0.2in}
\begin{quote}
    \textbf{Kevin}: Let's say someone was knocking on your door, right, and you said ``Who is it?'' to them but they didn't hear you or you kept saying it and they're trying to get in for some reason. You use the hologram so you could see who is outside using the cameras, and if there's somebody there you could speak to them using a mic or you could see what they're doing, or you could also contact 9-1-1.
\end{quote}
\endgroup

The value ``a comfortable life'' means a prosperous life and was reflected in 7 designs. This value is represented in both Ollie and Denny's smart light controllers intended to help them create a comfortable homework environment,

\begingroup
\addtolength\leftmargini{-0.2in}
\begin{quote}
    \textbf{Adult}: For you, Claude turns off the lights. If Adult owned Claude, could Adult train Claude to turn on music when it's homework time, or is Claude just a Denny thing that always turns off the lights?\\
    \textbf{Denny}: Adult could train him.\\
    \textbf{Adult}: Okay, so it could help anyone to make their homework environment no matter what, this is just your specific one? \{\textit{Denny nods}\}

\end{quote}
\endgroup

Alan's teachable machine reflected Rokeach's value ``inner harmony,'' which means to have freedom from inner conflict. Alan designed an emotion detection machine in order to identify and alleviate times when someone may be experiencing negative emotions like sadness or anger, as well as detect and reward times when someone is happy or excited.

Last, Penny's safety bracelet appealed to Rokeach's value, ``an exciting life,'' alternatively defined as a stimulating, active life. Penny drew inspiration from her extracurricular activity cheerleading involving difficult stunts and physical exertion. Penny describes how her teachable machine would detect cheerleader's quick actions and body movements,

\begingroup
\addtolength\leftmargini{-0.2in}
\begin{quote}
    \textbf{Penny}: It's gonna also alert when someone is falling or like some quick action or movement. When someone's falling it blows up the mat and tells people to get off the mat so that the faller can fall on the mat.
\end{quote}
\endgroup

\section{Discussion}
Our user study, exploratory in nature, shows both promising results and future research directions for teachable machines in problem formulation activities that aim to help children develop AI literacies. In this section, we first reflect on lessons learned: some offer new insights, others strengthen prior empirical and anecdotal evidence. We also discuss implications for designing AI problem formulation activities and broader use of design metaphors and human values as analytical lenses in AI problem formulation. We then discuss limitations in our study that may affect both the applicability of our approach and the generalizability of our findings.

\subsubsection*{Key aspects characterizing children’s formulated AI problems (RQ1).}

We find that children draw from their life experiences when identifying problems. In the home, their AI designs can provide security, control lights, automate cooking, fold clothes, discipline siblings, and call one’s mom for comfort. At school they solve math and puzzles, ensure safety in cheerleading, and bellow abuses to bullies. While the form factor of their designs varied, almost all assumed voice capabilities and many envisioned a state of constant video monitoring. Few restricted their designs to supervised classification, the machine learning paradigm in the teachable machines they were exposed to. Some moved away from classification to other supervised machine learning approaches such as learning by demonstration or imagined that their machines would learn in an unsupervised fashion, by just following them around and observing them. Others presumed no learning at all; their machines are ``\textit{smart}’’ and they just ``\textit{know}’’. The majority of those who chose to explicitly train their machines opted to provide labels via voice. Interesting concepts like synthetic data, discriminatory features, similarity, and fine-grained classification emerged in few of the designs.  Most of the children anticipated errors. They were related to false positives, false negatives, grasping and object handling, force and torque control, and latency. Many of them would fix the errors by providing more training data, two opted to restart the training, and one imagined that the machine ``\textit{can debug itself so there isn’t anything to worry about.}''

\subsubsection*{Prevalent design metaphors for AI in children’s ideas (RQ2).} We find that when children are first asked to train a machine and then build their own to solve a problem of their choice, they mostly do not anthropomorphize their machines beyond naming them or giving them pronouns.
Instead, children's designs tend to align with proposed changes in design metaphors for AI~\cite{shneiderman2022human, chan2021supporting}; they moved away from \textit{intelligent agents}, \textit{teammates}, \textit{assured autonomy}, and \textit{social robots}.
The most prevalent design metaphors reflected in children's ideas were \textit{supertools}, \textit{tele-bots}, \textit{control centers}, and \textit{active appliances}.
Shneiderman suggests these same metaphors for designers to pursue, \ie, ``successful designers avoid mimicking human models and pursue supertools, tele-bots, active appliances, and control centers that support human control over technology while ensuring high levels of automation.''
Such an alignment concurs with what Shneiderman calls the \lineQuote{Innovation Goal} that ``drives researchers to develop widely used products and services by applying HCAI methods.''

\subsubsection*{Personal values incorporated into children's designs (RQ3).} Ongoing discourse on human-centered AI routinely advocates for considering values such as reliability, safety, and trustworthiness in future AI technology. 
We found that children’s designs for teachable machines often reflected similar values including ``Responsibility,'' ``Capability,'' and ``Family Security.'' 
Furthermore, many of the children's designs reflected the value ``Obedience,'' which draws analogy to Shneiderman’s~\cite{shneiderman2022human} design metaphors that stress extending human abilities such as human control.
However, we also found more nuance in terminal values related to security and safety. 
Current discourse around safety and security in AI tends to emphasize misuse, potential weaponization or ethically questionable goals \eg hacking or gaming a system loophole to complete a task~\cite{brief2019ai}. 
While the children's designs mapping to the terminal goal of ``family security'' echoed these discussions, they also expanded the value space to consider comfort, familial connection and harmony, along with excitement \eg keeping acrobatic cheerleaders safe.
These values for family and comfort were high priorities for children; yet nuances around comfort and harmony are often overlooked in existing AI design considerations, ethical debates, or frameworks.

\subsection{Implications}

\subsubsection{Youth \& AI Literacies.} We provide empirical evidence on the affordances of participatory design activities with teachable machines for engaging children, as young as 8 years old, in AI problem formulation. We demonstrate that established co-design approaches~\cite{druin1999cooperative, fails2012methods, guha2013cooperative} with problem reduction heuristics~\cite{maccrimmon1976decision} in the context of teachable machines can promote children's AI literacies and support their efforts to formulate problems for AI-infused technologies with personal relevance in their daily lives. Despite their inexperience with machine learning, the children worked closely with adults to explore what was feasible and what was not, iteratively reformulating the problem space as they considered their machine's specific failure and recovery conditions. Their designs showcase their creativity not only in the personally meaningful problems they articulated, but also in how they tried to construct their training strategies.

We situate our work in Long and Magerko's~\cite{long2020ailiteracy} definition of AI literacy and connect our study with several of their design considerations, along with competency goals that our sessions fulfill. 
Since we based our sessions on teachable machines that don't require coding and used pen and paper-based storyboarding, our sessions exemplify Long and Magerko's design considerations for \textit{Low Barrier to Entry}.
By structuring our storyboard and asking children questions that promote reflection on their systems, we promote the design considerations on \textit{Critical Thinking}. Finally, children were free to solve a problem of their choice. We found personally meaningful ideas like Penny's safety bracelet for her cheerleader team, which reflected their design consideration of \textit{Identity, Values, and Backgrounds}.

As AI and machine learning technology proliferate in children's everyday lives, we gain more opportunities to equip them better to tackle overhyped claims about the infallibility of AI.
To this end, our AI problem formulation activity fulfill many competency goals from Long and Magerko's~\cite{long2020ailiteracy} work 5 that intersect with the learning goals laid out by Touretsky~\cite{touretzky2019learningactivities}.
Empowering children to make a teachable machine ensures that they are not reduced to mere ``users'' of AI. 
Our work shows how children ground their ideas in the ``data'' used to train a machine. 
Hence, our approach to using teachable machines, similar to Dwivedi~\etal~\cite{dwivedi2021introducing} and Vartiainen~\etal~\cite{vartiainen2021machine} followed by a structured storyboarding session can be beneficial in fulfilling the competency goal of \textit{Learning from Data}.
We saw that the storyboard's scaffolded structure helped children reflect on different aspects of their designs.
In combination with recent work by Vartiainen~\etal~\cite{vartiainen2021machine} and Hitron~\etal~\cite{hitron2018introducing}, our storyboarding activity, which focuses on potential errors and options for error recovery, demonstrates how we can encourage children to reflect critically on the utility and limits of AI in their everyday lives and fulfill the competency goal of A\textit{I's Strengths and Weaknesses.
Finally, the children design their machines and formulate the machine learning problems they want to solve by imagining their own AI and fulfilling the competency goal to \textit{Imagine Future AI}.
}
\begin{figure*}[t]
     \begin{tabularx}{\textwidth}{XX}
    \centering
    \begin{subfigure}[b]{.43\textwidth}
         \includegraphics[width=\linewidth]{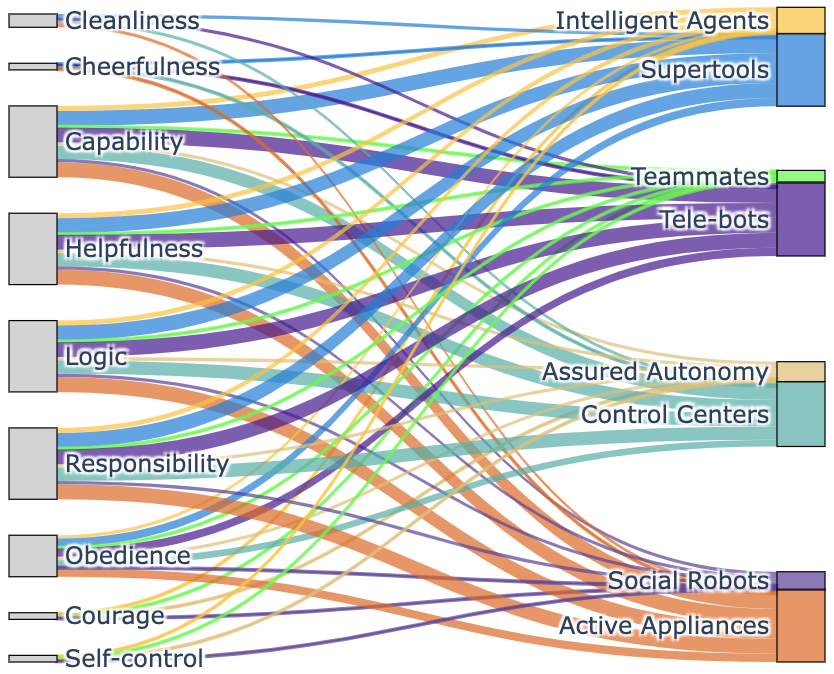}
         \caption{Instrumental Values \& Design Metaphors}
         \label{fig:IV_DM}
     \end{subfigure}
     &
    \centering
    \begin{subfigure}[b]{.43\textwidth}
         \includegraphics[width=\linewidth]{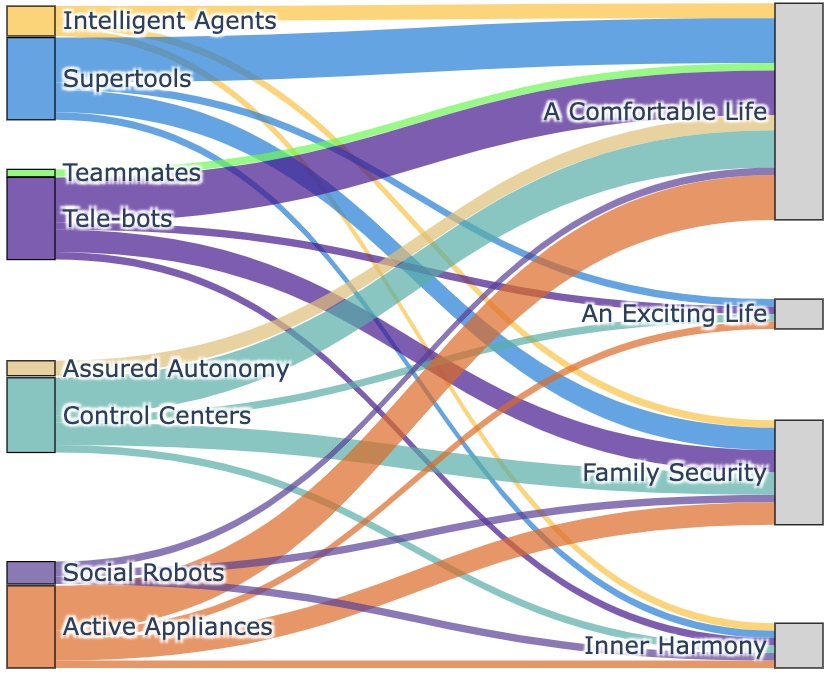}
         \caption{Design Metaphors \& Terminal Values}
         \label{fig:DM_TV}
     \end{subfigure}
         \end{tabularx}
        \caption{These Sankey diagrams illustrate the relations between the design metaphors and human values in the children's designs. Each flow line represent the strength of a link (or affinity) between the two \eg, \lineQuote{Active Appliances} are closer to Instrumental Values like ``Helpfulness,'' ``Capability'', ``Logic'' and ``Responsibilty''; and the Terminal Values of ``A Comfortable Life'' and ``Family Security.'' }
        \label{fig:valuesvsdm}
\end{figure*}

\subsubsection{Youth \& AI Ethics}
Gaining insight into children's values has been a longstanding challenge in Child-Computer Interaction (CCI)~\cite{elsayed-ali2020designing,spiel2018micro-ethics,van_mechelen2014applying,skovbjerg2016being,druin2002role,iversen2012values-led,iversen2012working}, and attending to children's values can promote critical computational action practices~\cite{tissenbaum2019computationalaction} and positive dispositions toward STEM.  
Our study demonstrates that when children and adults co-design together, they engage in and enjoy critical reflection, playfully questioning the boundaries of privacy and AI ethics issues.
For example, children asked their machines to cheat on math quizzes to help reduce math anxiety, to read others' minds as a means for nurturing friendships and mitigating social anxiety, and to serve as powerful physical and verbal allies against bullying.
These discussions indicate that storyboarding approaches like ours may promote reflexive conversations around the ethics of AI model development and deployment.
Similarly, Williams~\etal~\cite{williams2023ai} and Zhang~\etal~\cite{zhang2023integrating} showed that middle schoolers brought up ethics when designing their own AI systems that included image recognition and generation. 
Another recent example comes from Kusuma \etal~\cite{kusuma2022civilwar} involving adults who explored ethical considerations in facial recognition for finding relatives in Civil War databases.
Future researchers could extend the approaches from this prior work with a storyboarding session like ours that includes prompts for participants to fix their systems' ethical issues. 
We show that when  grappling with technical AI issues related to personally meaningful topics, children can be creatively involved and immersed in problem formulation and critical of potential ethical limits of AI.
We hope that our choice of widely available resources and materials could encourage others to replicate and expand on this work to facilitate reflections on AI ethics and engage children from other developed or developing regions and across cultures. For instance, one could combine the idea of a pre-defined toolkit from Druga~\etal~\cite{druga2022family} and our structured storyboarding activity to further ground the session within machine learning concepts.

\subsubsection{UX \& AI Practitioners.} There is an increasing need among practitioners for support during the early stages of ideation and problem formulation in order to preempt AI product failures~\cite{yildirim2023investigating} with relevat skills being foregrounded as foundational, enduring, and adaptable competencies~\cite{acar2023ai}. 
We see how insights from this study could contribute to future research with this broader population. For example, we observed that after engaging with a teachable machine, children would be critical of AI if asked to find out how their machines would fail, did little anthropomorphizing of their machines, and aligned more closely with Shneideman's call for a shift in design metaphors for AI~\cite{shneiderman2022human, chan2021supporting} despite influence by existing technologies and science fiction. This observation strengthens findings in prior work by Druga et al~\cite{druga2021how} of children's increased awareness of the potential fallibility of AI systems, after they experimented with teachable machines.
However, the children's shifts toward more nuanced, less techno-utopian views also lead us to reflect whether activities with teachable machines, further enriched with our analytical lenses, could also be structured for meaningful engagement of UX practitioners, who may not have machine learning expertise, in AI problem formulation. 
For instance, after interacting with teachable machines, UX practitioners could storyboard with other stakeholders imagined AI systems, map the ideas to Shneiderman’s design metaphors and Rokeach's value framework, reflect, and iterate.
Such activities would build upon prior value-sensitive design by Shilton~\cite{shilton2013valueslevers} that established ``practices that open new conversations about social values and encourage consensus around those values as design criteria''~\cite[p.~374]{shilton2013valueslevers}.

Another example could be future research focusing on AI practitioners.
Machine learning researchers are taking AI-centric approaches to label human values, borrowing from deontological or moral values instead of material values~\cite{emelin2020moral, pan2023rewards}.
However, emerging work in value alignment of AI~\cite{han2022aligning} (such that it benefits human society) suggests using material values to guide the ``actions of AI agents that are preferable to other actions'' because material values are realized through established human value frameworks like the Rokeach Value Survey~\cite{rokeach1973nature}. 
We can understand this tension by highlighting the relationship between design metaphors and human values: the same tension exists between ``Innovation Goals'' and ``Science Goals.''
We intentionally use the word ``lens'' when describing problem formulation. By combining design metaphors, or value lenses, we gain different perspectives of the design space.
We can see Figure ~\ref{fig:valuesvsdm} as a reflection of this design space within children's ideas.
When using design metaphors to guide the design robots~\cite{dennler2022using}, designers could use Rokeach Values to understand what human values it reflects.
For example, we see that children's ideas with ``Control Centers'' and ``Tele-bots'' are related to human values of ``Responsibility'' and ``Obedience.''
And so if a robot is designed in contravention of this design space, say, as an ``autonomous teammate'' but one which fails to communicate that it is ``obedient'' to the human, we can expect such a system to fail because a user would not feel in control of the robot.
Our work may inspire future research into how to weave human values into design metaphors and how AI practitioners can leverage human value frameworks such as Rokeach's in their optimization objectives.

\subsection{Limitations}
Our co-design study is exploratory and subjective in nature, thus it is not conclusive. It provides a rich set of observations and insights, generating hypotheses that need further investigation. For instance, to demonstrate the effectiveness of our storyboarding activity over other approaches with teachable machines, such as sketching, prototyping, and brainstorming,  a mixed-methods approach is needed. Similarly, demonstrating effectiveness on learning outcomes would require pre-post measurements.  

\subsubsection{One-to-one child-adult ratio.}
First, while all participating children had experience training and testing a teachable machine, we had a one-to-one ratio of adult to children co-designers.
Such a ratio may not be possible in other AI literacies spaces, such as in a formal classroom where it is likely that only one adult can facilitate.  
In addition, future studies should explore how problem formulation discussions unfold when there are more children than adults and in a formal learning context.

\subsubsection{Acknowledging child development differences.}
Second, the children participating in this study's problem formulation activity ranged in age from 8-13 years old, and we acknowledge child development differences across this six-year time span. The small number of child partners (n=10), while representative of sample sizes in user studies in human-computer interaction~\cite{caine2016local}, did not allow us to substantiate or unpack potential effects of child development. 
However, as noted above, all the children had experience testing and training teachable machines (figure \ref{fig:sessions}, session 1), so they all had similarly grounded exposure to machine teaching concepts. 
In addition, all the older children (11-13) and only one 8-year old (Brian) took part in session 2 (the museum-based machine-teaching experience). 
Anecdotally, we also observed age-related patterns in the children's creations: despite working more closely with the older children, 8-year old Brian's final machine revealed similar values, design metaphors, and overall functionality as the other 8-year old boys.

\subsubsection{Limited accessibility.}
Third, we constructed our printed storyboards and selected other session materials like videos that were included in presentation slides for this activity with children that we knew apriori would be sighted. While both the storyboard and the videos could be made accessible via a screen reader (\eg, convert the storyboard to an accessible slide and provide audio description for the videos), more work is needed to explore equally effective co-design sessions especially for mixed-abilities teams and groups. 

\section{Conclusion}
Recent efforts in participatory machine learning~\cite{sloane2020participation, birhane2022frameworks, bondi2021envisioning} call for designers and researchers to carefully consider what we intend by ``meaningful participation'' and who we involve as partners in our design work. 
Our work responds to these calls by preparing children -- those who may be the most affected by increasingly pervasive AI technologies -- as partners in this process. 
The goal of this study was to expand the level of control and creativity children can exercise with teachable machines by engaging them in a specific problem formulation exercise.
To pursue this goal, we conducted a modified storyboarding session with youth 8-13 years old.
Our findings and exploratory insights contribute to the design of learning activities that use teachable machines.
Particularly, they could benefit from allowing children to formulate their machine learning problems, using children's values to be usable and enjoyable, and showcasing their utility to support children's goals.
We call on future designers and researchers to conduct more studies that involve children as active agents in the design of everyday AI systems imbued with their values.

\begin{acks}
We thank the KidsTeam children for their ideas, input, and contributions.
We also thank graduate students Merijke Coenraad, Nitzan Koren, Zahra Zuzar Dhuliawala, Pooja Bipin Gajera, Naishi Jain, Dinesh Kumar Nanduri, and undergraduate students Anumta Ali and Subhatra Sivam for their contributions in conducting the study sessions as adult co-designers of KidsTeam.
Utkarsh Dwivedi was partially supported by the Ann G. Wylie Dissertation Fellowship and NSF grant \#1816380.
Hernisa Kacorri was partially supported by NSF grants \#1816380, \#1955568, and \#2229885. 
\end{acks}
\bibliographystyle{ACM-Reference-Format}
\bibliography{bib}

\end{document}